\documentclass[
    ,final            % use final for the camera ready runs
  ]
  {aipproc}

\layoutstyle{6x9}

\usepackage{natbib}
\usepackage[english]{babel}
\usepackage{epsfig}
\usepackage{psfig}
\usepackage{graphicx}

\begin{document}

\title[Transport in self-gravitating accretion discs]{Testing the
  locality of transport in self-gravitating accretion discs}

\author{Giuseppe Lodato}{
  address={Insitute of Astronomy, Cambridge, UK}
}

\author{W. K. M. Rice}{
  address={School of Physics and Astronomy, University of St. Andrews,
St. Andrews, UK} }

\begin{abstract}
In this paper we examine the issue of characterizing the transport
associated with gravitational instabilities in relatively cold discs,
discussing in particular under which condition it can be described
within a local, viscous framework. We present the results of global,
three-dimensional, SPH simulations of self-gravitating accretion
discs, in which the disc is cooled using a simple parameterization for
the cooling function. Our simulations show that the disc settles in a
``self-regulated'' state, where the axisymmetric stability parameter
$Q\approx 1$ and where transport and energy dissipation are dominated
by self-gravity. We have computed the gravitational stress tensor and
compared our results with expectations based on a local theory of
transport. We find that, for disc masses smaller than $0.25M_{\star}$
and aspect ratio $H/r<0.1$, transport is determined locally,
thus allowing for a viscous treatment of the disc evolution.
\end{abstract}

\maketitle

%%%%%%%%%%%%%%%%%%%%%%%%%%%%%%%%%%%%%%%%%%%%
%% MAINMATTER
%%%%%%%%%%%%%%%%%%%%%%%%%%%%%%%%%%%%%%%%%%%%

\section{Introduction}

One of the basic unknowns in accretion disc theory is the physical
mechanism ultimately responsible for angular momentum transport and
energy dissipation in the disc. The usual way to overcome this
difficulty is to assume that transport is dominated by some kind of
viscous process, and to give an {\it ad hoc} prescription for the
$r\phi$ component of the stress tensor $T$ (usually vertically
integrated, under the assumption that the disc is geometrically
thin). The most widely used prescription is the so-called
$\alpha$-prescription \citep{shakura73}, according to which:
\begin{equation}
T_{r\phi}=\left|\frac{\mbox{d}\ln\Omega}{\mbox{d}\ln
  r}\right|\alpha\Sigma c_s^2,
\label{alpha}
\end{equation}
where $\Omega$ is the angular velocity of the disc, $c_s$ is the
thermal speed, $\Sigma$ is the surface density, and $\alpha$ a
free parameter, expected to be smaller than unity. 

It has recently been recognized that accretion discs threaded by a
weak magnetic field are subject to MHD instabilities (see
\citet{balbusreview} and references therein), that can induce
turbulence in the disc, thereby being able to transport angular
momentum outwards and to promote the accretion process. However, in
many astrophysically interesting cases, such as the outer regions of
protostellar discs, the ionization level is expected to be very low,
reducing significantly the effects of magnetic fields in determining
the dynamics of the disc. A possible alternative source of transport
in cold discs is provided by gravitational instabilities.

The axisymmetric stability of a thin disc with respect to
gravitational disturbances is determined by the condition
\citep{toomre64}:
\begin{equation}
Q=\frac{c_s\kappa}{\pi G \Sigma}>\bar{Q}\approx 1,
\label{Q}
\end{equation}
where $\kappa$ is the epicyclic frequency. The external regions of
many observed systems are likely to be gravitationally unstable. If we
consider, for example, the case of AGN accretion discs, it can be
shown \citep{LB03a} that $Q$ falls below unity already at a distance
of $\approx 10^{-2}$ pc from the central black hole. In the case of
protostellar discs, models of the outburst in FU Orionis systems (a
class of young stellar objects that experience a phase of enhanced
accretion) show that the disc is marginally stable already at a
distance of $\approx 1$ AU from the central young star
\citep{bellin94}.

It has been suggested \citep{pacinski78,BL99} that, in a cold enough
disc, where condition (\ref{Q}) is violated, gravitational
instabilities would provide a source of energy dissipation, therefore
heating up the disc and leading to a self-regulated state, with the
disc close to marginal stability. Gravitationally induced spiral
structures are also able to transport angular momentum. In this paper
we examine the issue of characterizing the angular momentum transport
and energy dissipation in self-gravitating discs. In particular, we
would like to test the extent to which the transport due to
gravitational instabilities can be described as a local, viscous
process, amenable to a description analogous to Eq. (\ref{alpha}), as
proposed by \citet{lin87}.

\citet{balbus99} have argued that the long-range nature of the
gravitational field would preclude a local description of transport,
as implicitly assumed in a viscous scenario. In particular, they have
shown how the energy flux contains extra terms, related to wave
transport, which are ``anomalous'' with respect to a viscous
flux. \citet{gammie01} has performed local, shearing sheet,
zero-thickness simulations of self-gravitating accretion discs, and
has concluded that a local description is adequate in such
``razor-thin'' discs, extrapolating his results to discs for which
$H/r< 0.1$. However, Gammie's simulations are not appropriate
to test global effects, since locality is set up from the beginning,
and they are only valid for infinitesimally thin
discs. \citet{rice03a,rice03b} have already shown using global, 3D
simulations, how global effects can be important in the dynamics of
self-gravitating discs, with respect to the related issue of
fragmentation. Here we follow up the work of Rice et al., in order to
quantify under which conditions a local model for transport in
self-gravitating discs is justified.

\section{Numerical setup}

The simulations presented here were performed using smoothed particle
hydrodynamics (SPH), a Lagrangian hydrodynamics code
\citep[see][]{monaghan92}. Our code uses a tree structure to calculate
gravitational forces and the nearest neighbours of particles. The
particles are advanced with individual time-steps \citep{bate95},
resulting in an enormous saving in computational times when a large
range of dynamical timescales are involved.

We consider a system comprising a central star, modelled as a point
mass $M_{\star}$, surrounded by a disc with mass $M_{disc}$. We have
performed several simulation using different values of the parameter
$q=M_{disc}/M_{\star}$. The initial surface density profile was taken
to be a power-law $\Sigma\propto r^{-1}$. The disc is initially
gravitationally stable and the minimum value of $Q$ is $Q_{min}=2$. 
Our calculations are essentially scale-free. In dimensionless units,
the disc extends from $r_{in}=0.25$ to $r_{out}=25$. We have used
$N=250000$ particles.

We use an adiabatic equation of state, with adiabatic index
$\gamma=5/3$. We explicitly solve the energy equation for our system,
including heating from shocks and from artificial viscosity and
introducing a cooling term, simply parameterized as:
\begin{equation}
\left(\frac{\mbox{d}u_i}{\mbox{d}t}\right)_{cool}=-\frac{u_i}{t_{cool}},
\end{equation}
where $u_i$ is the internal energy of a representative particle, and,
as in \citet{gammie01} and in \citet{rice03b}, the cooling time-scale
$t_{cool}=\beta\Omega^{-1}$, where $\beta$ is a free parameter,
independent of radius. \citet{gammie01} and \citet{rice03b} have shown
that if cooling is too fast, then the disc will fragment as a result
of the gravitational instabilities. In our simulations we have used
$\beta=7.5$, in which case none of our discs was found to fragment.

\section{Results}

We have run our simulations for a few thermal timescales, in order to
reach thermal equilibrium.  In all cases the disc initially cools down
until the value of $Q$ becomes small enough for gravitational
instabilities to become effective and to provide a source of heating
to the disc. At later stages the disc settles into a self-regulated
state with $Q\approx 1$ over a large portion of the disc.

\begin{figure}
\centerline{ \epsfig{figure=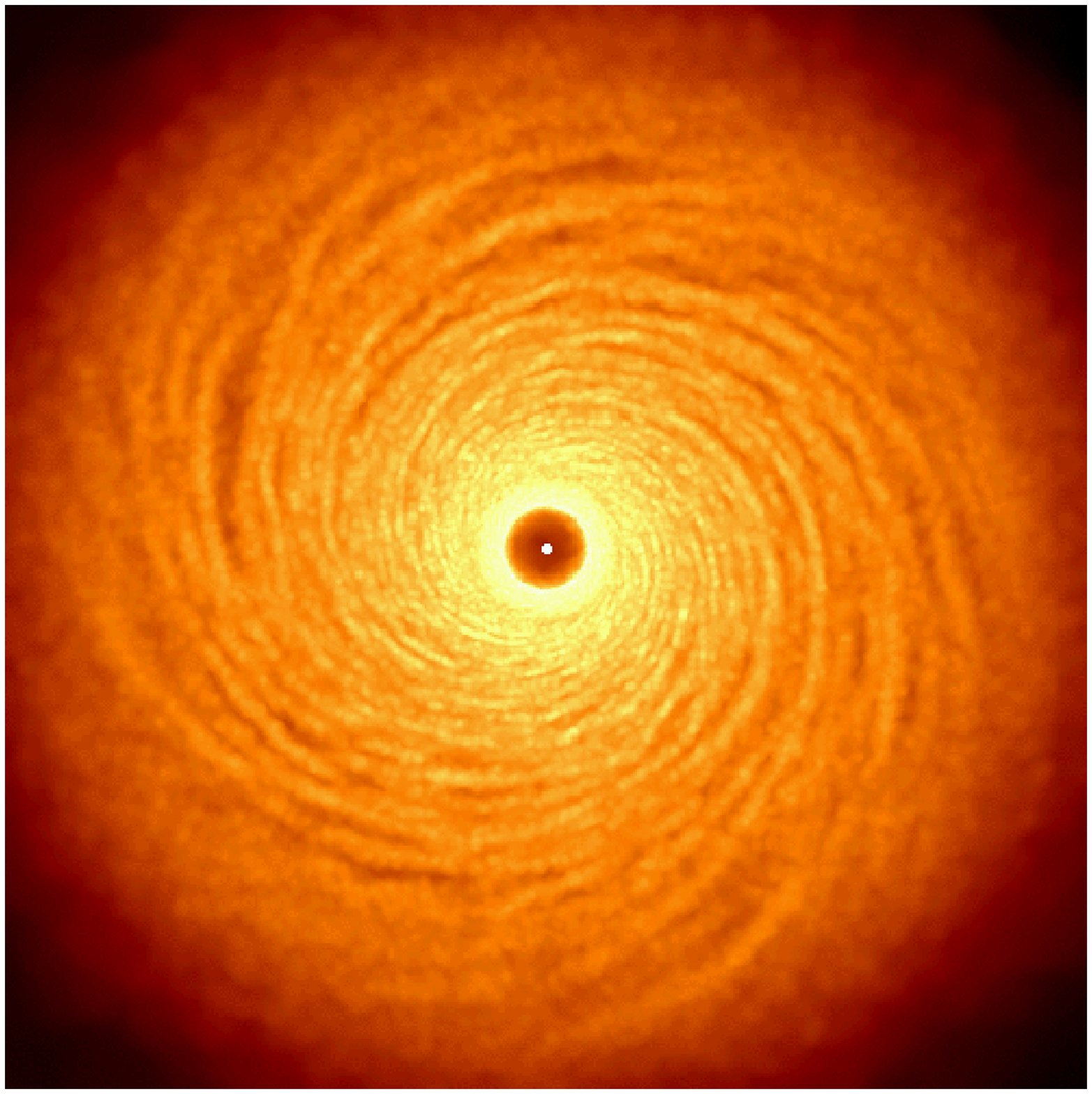,height=.27\textheight}
             \epsfig{figure=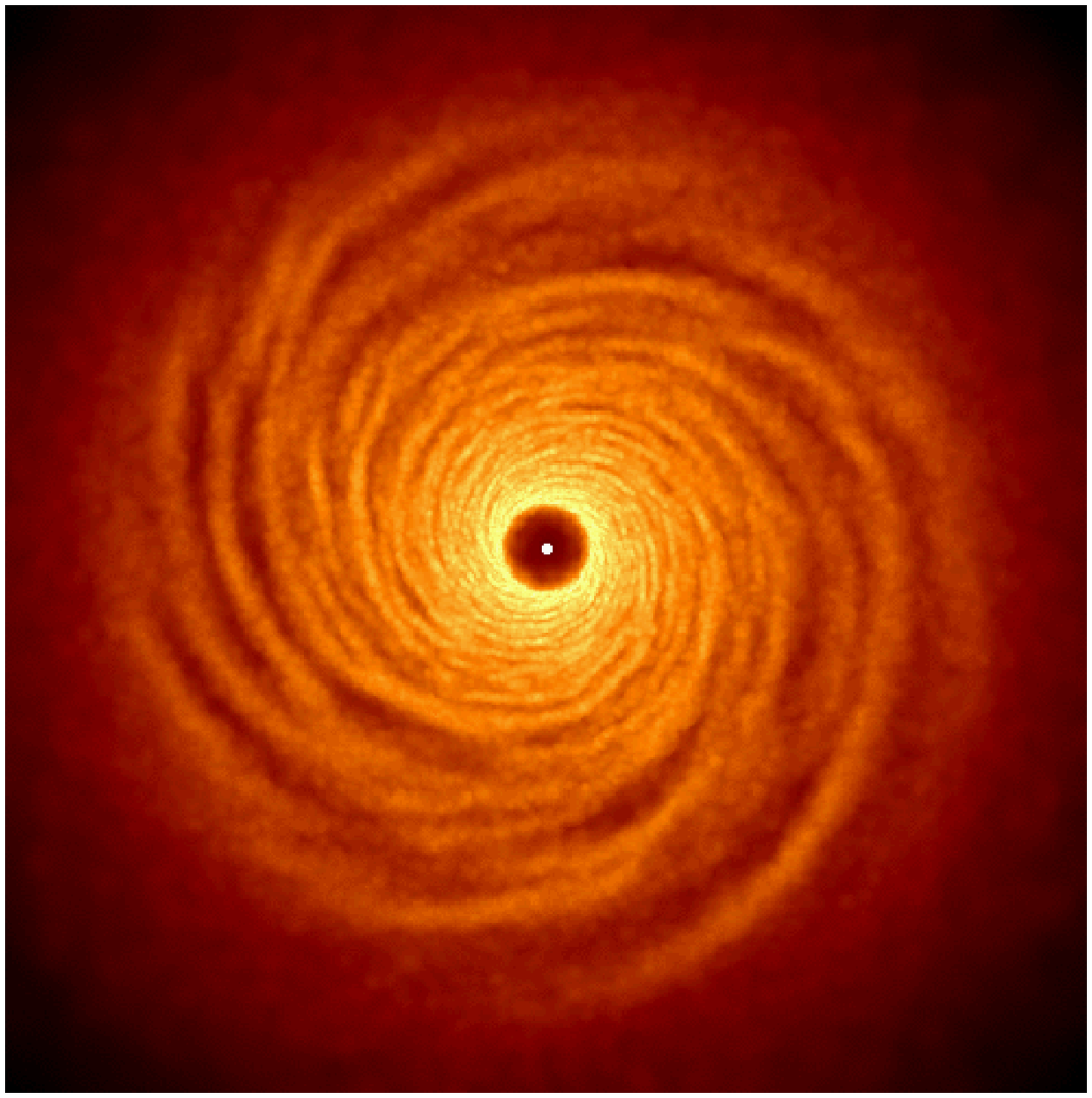,height=.27\textheight}
	     \epsfig{figure=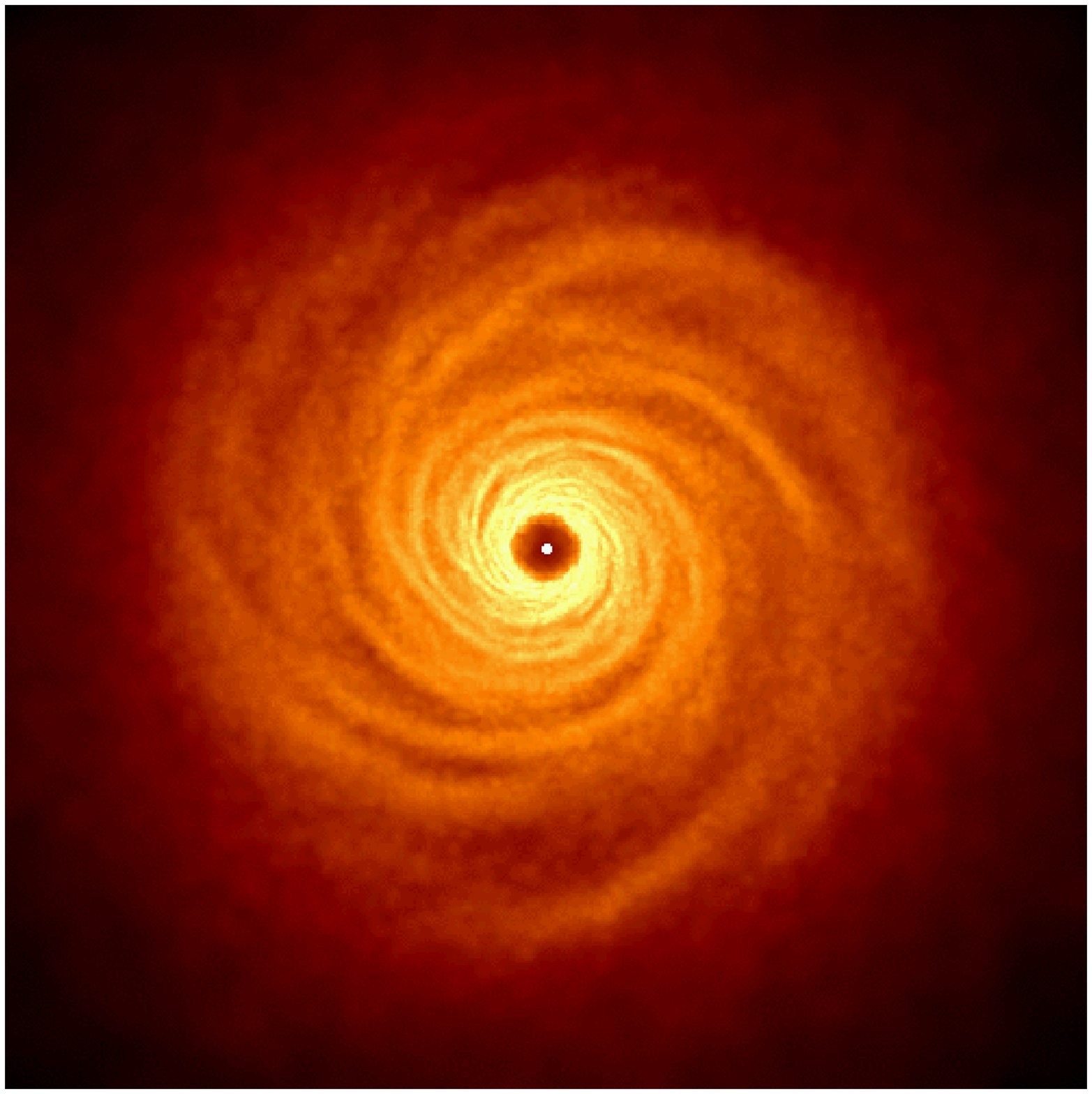,height=.27\textheight}}
  \caption{Equatorial density structure at the end of the simulations
  for (left) $q=0.05$, (middle) $q=0.1$, and (right) $q=0.25$.}
\label{fig2}
\end{figure}

\begin{figure}
\centerline{ \epsfig{figure=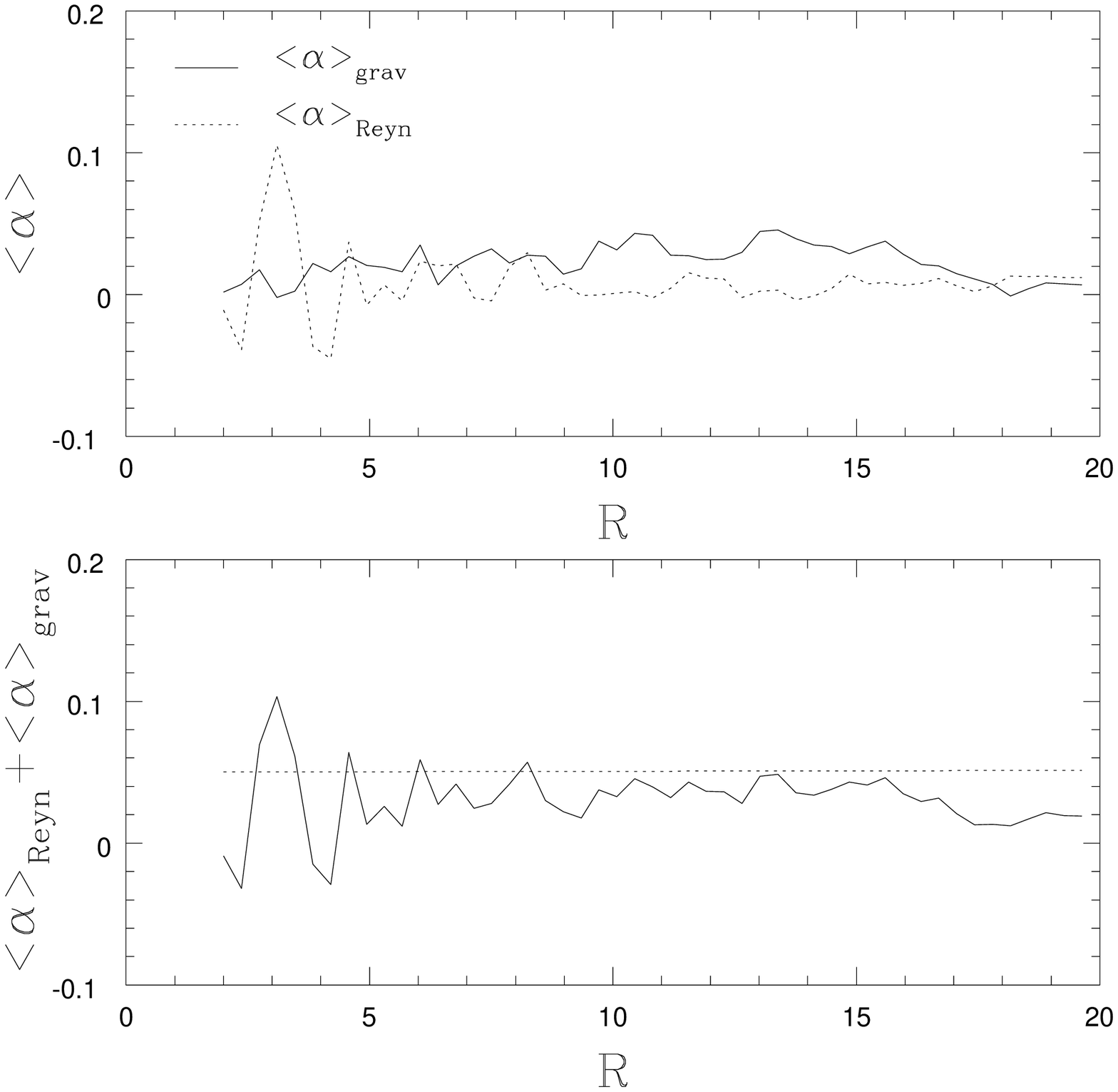,height=.28\textheight}
             \epsfig{figure=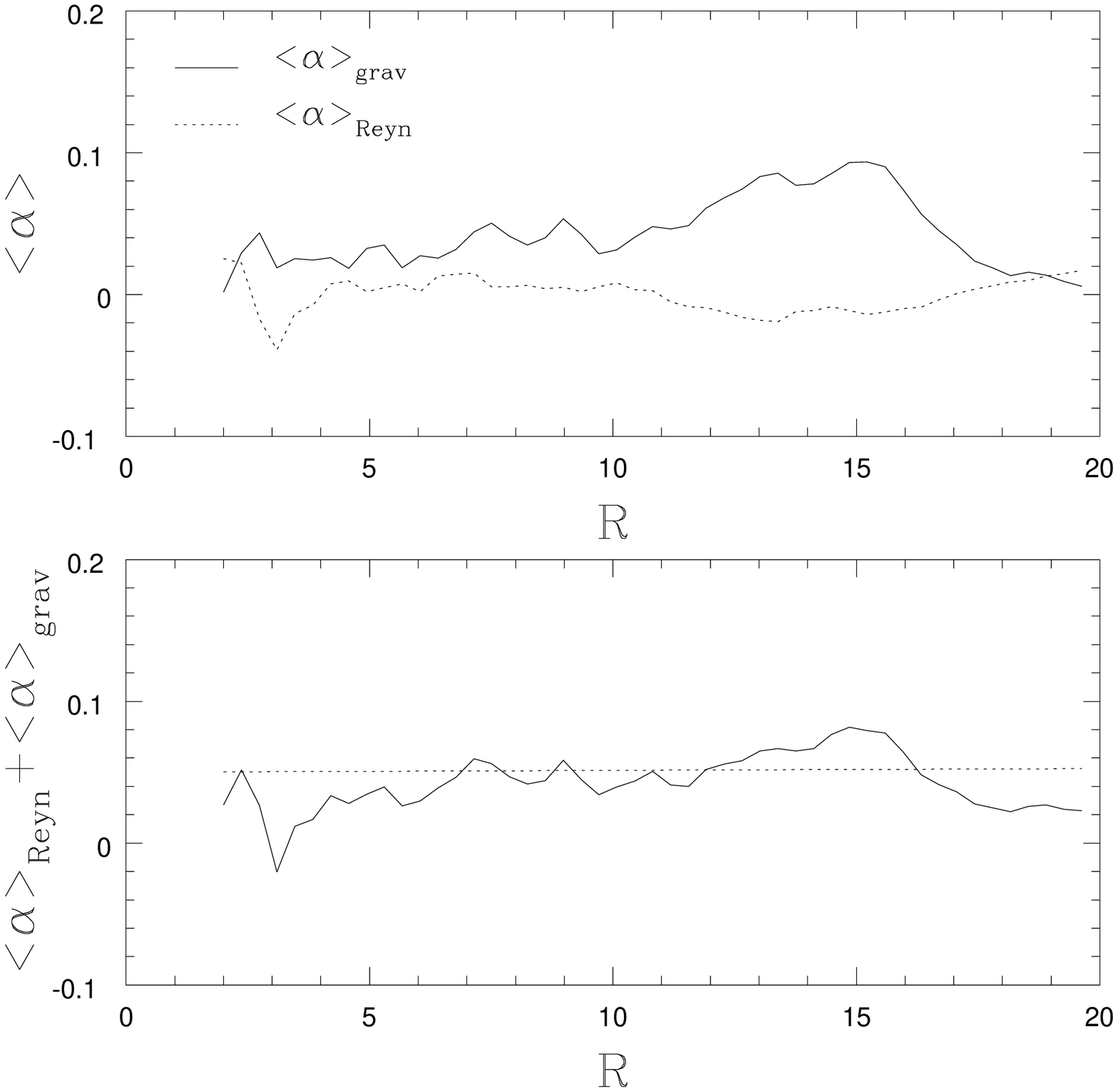,height=.28\textheight}
	     \epsfig{figure=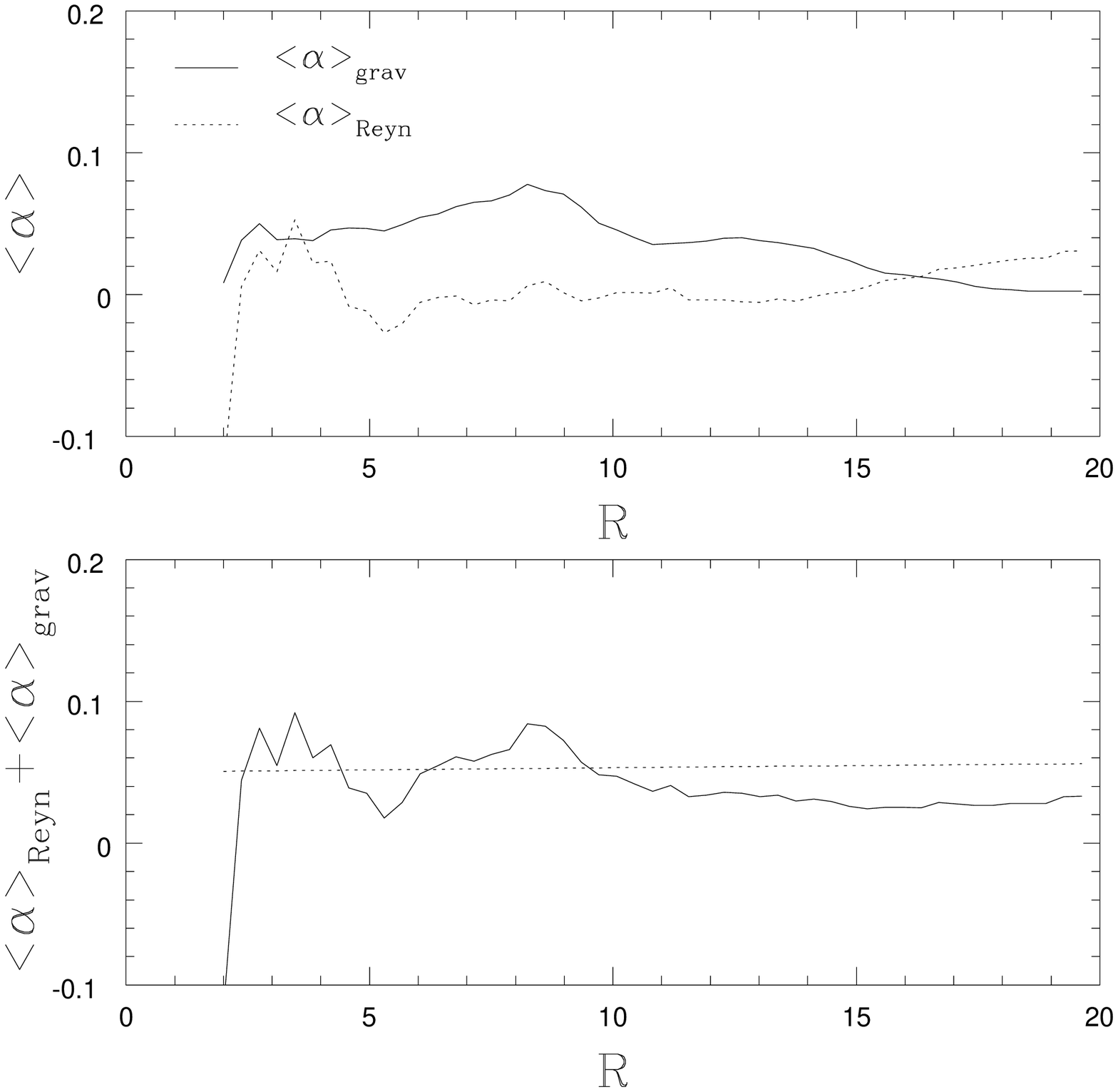,height=.28\textheight}}
 \caption{Effective $\alpha$ produced by gravitational instabilities
 for (left) $q=0.05$, (middle) $q=0.1$, and (right) $q=0.25$. The top
 panel shows the separate contribution of $\alpha_{grav}$ and
 $\alpha_{Reyn}$, the lower panel shows the sum of the two
 contributions compared with the expected value from a local viscous
 model (dotted line).}
\label{fig1}
\end{figure}

The $r\phi$ component of the stress tensor (integrated in the
$z$-direction) associated with self-gravity is given by
\citep{lyndenbell72}:
\begin{equation}
T_{r\phi}^{grav}=\int\mbox{d}z\frac{g_rg_{\phi}}{4\pi G},
\label{stress}
\end{equation}
where $\bf{g}$ is the gravitational field of the disc. The hydrodynamic
(Reynolds) contribution to the stress tensor is given by:
\begin{equation}
T_{r\phi}^{Reyn}=\Sigma {\delta v}_r{\delta v}_{\phi},
\label{hydro}
\end{equation}
where ${\bf{\delta v}=\bf{v}-\bf{u}}$ is the velocity
fluctuation, $\bf{v}$ is the fluid velocity and $\bf{u}$ is
the mean fluid velocity. 

After averaging the stress tensor azimuthally and radially, over a
small region $\Delta r=0.1r$, we have computed the corresponding value
of $\alpha$ (see Eq. (\ref{alpha})):
\begin{equation}
\alpha(r)=\left|\frac{\mbox{d}\ln\Omega}{\mbox{d}\ln
r}\right|^{-1}\frac{<T_{r\phi}^{grav}>+<T_{r\phi}^{Reyn}>}{\Sigma c_s^2}.
\end{equation} 

The resulting radial profiles of $\alpha$ are shown in Fig. \ref{fig1}
for the three cases $q=0.05$, $q=0.1$, and $q=0.25$. The upper panels
show separately the hydrodynamic and gravitational contributions to
$\alpha$, while the bottom panels show the sum of the two. The results
are also averaged in time, over the last 500 timesteps of the
simulations. 

The simple parameterization of the cooling time adopted here allows us
to have a direct insight on the problem of characterizing transport
properties and dissipation in the disc, by using the following result
\citep{pringle81}, which follows directly from the energy balance
equation: if the disc is in thermal equilibrium, and {\it if heating
is dominated by a purely viscous process}, described by
Eq. (\ref{alpha}), then the viscosity coefficient $\alpha$ and the
cooling time-scale $t_{cool}$ satisfy the following relation:
\begin{equation}
\alpha=\left|\frac{\mbox{d}\ln\Omega}{\mbox{d}\ln
r}\right|^{-2}\frac{1}{\gamma(\gamma-1)}\frac{1}{t_{cool}\Omega}
\label{alphaexp}
\end{equation}

In our simulations $t_{cool}=7.5\Omega^{-1}$, so from
Eq. (\ref{alphaexp}) we would expect that in thermal equilibrium, if
energy dissipation can be described locally, $\alpha$ should be
approximately constant with radius (this is indicated with the dotted
line in the bottom panels of Fig. \ref{fig1}). We can interpret the
bottom panels of Fig. \ref{fig1} in the following way: the solid line
gives a measure of the actual torque induced by gravitational
instabilities in our simulations, while the dotted line is the torque
that a viscous process should exert in order to dissipate the amount
of energy which is actually dissipated in the simulations. Therefore,
when the computed value of $\alpha$ is larger than the expected one,
gravitational instabilities dissipate {\it less} energy than a viscous
process, whereas in the opposite case, they dissipate {\it more}
energy than a viscous process.  Our results indicate that, up to a
disc mass $M_{disc}=0.25M_{\star}$, gravitationally induced transport
is fairly well described within a local framework. This result could
also be anticipated, based on Fig. \ref{fig2}, that shows how the disc
dynamics is dominated by high-$m$ modes, that dissipate on a short
length-scale.

As a separate test for the locality of transport, we have also
computed $\alpha_{part}(r,d)$, which is defined as the gravitational
part of $\alpha(r)$, where the gravitational field ${\mathbf g}$ is
computed taking into account only those particles which are inside a
spherical radius $d$ from the radial point $r$ at which we are
computing the stress. This quantity gives us a measure of the size of
the region that mostly contributes to the stress at a given point.

\begin{figure}
\centerline{ \epsfig{figure=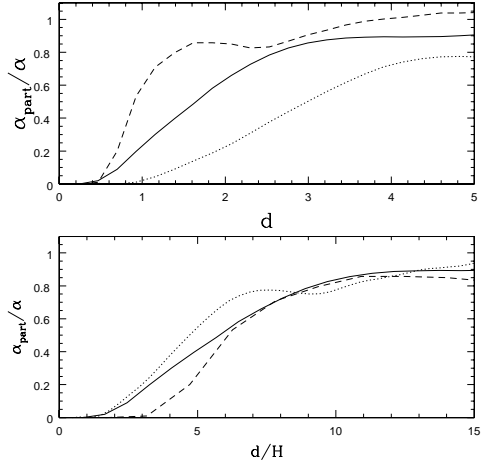,height=.3\textheight}}
 \caption{Contribution to the stress at $r=15$ from regions of the discs
 at a distance $d$ from $r$. Dashed line: $q=0.05$, solid line:
 $q=0.1$. Dotted line: $q=0.25$}
\label{fig3}
\end{figure}

Fig. \ref{fig3} shows the results for $r=15$ for the three
simulations. The upper panel shows clearly that the more massive the
disc is, the more far regions contribute to the stress. However, it
should be noted that, since $Q\approx 1$, a more massive disc is also
hotter, and therefore thicker. The bottom panel of Fig. \ref{fig3}
shows $\alpha_{part}$ as a function of $d/H$, where $H$ is the disc
thickness. In all cases, more than 80\% of the contribution to the
gravitational torque come from a region with size $\Delta\approx
10H$. Transport is going to be local if $\Delta\ll r$. We can
therefore conclude that, as long as $M_{disc}< 0.25M_{\star}$,
a disc with $H/r \ll 0.1$ will be reasonably described in terms of a
local model. This is basically the global analogous of the results
obtained by \citet{gammie01} based on a local, 2D model.

Note that detailed vertical structure models of FU Orionis objects
\citep{bellin94} lead to disc thickness $H/r\approx 0.1$, so that in
this case we could expect global effects to be important in
determining the emission properties of the disc, as suggested by
\citet{LB2001}.

\section{Conclusions}

We have performed global, three-dimensional SPH simulations of
self-gravitating accretion discs. Our results show how the heating
provided by gravitational instabilities can balance the cooling (that
we have parameterized in a simplified way), leading to a self-regulated
state where $Q\approx 1$. With a cooling rate
$t_{cool}=7.5\Omega^{-1}$, none of our simulations was found to
fragment. We have characterized the transport properties of the disc,
by computing the stress tensor associated with gravitational
instabilities and comparing the corresponding viscous dissipation rate
to the actual heating rate in the disc. We have found that, if the
disc is less massive than $0.25M_{\star}$, gravitational disturbances
are dominated by high-$m$ modes, so that the transport properties of
the disc can be described reasonably well in terms of a local, viscous
formalism. In particular, we have found that the gravitational torque
comes from a region of size $\approx 10H$, where $H$ is the disc
thickness. We can therefore expect that discs with an aspect ratio
$H/r< 0.1$ can be treated locally.

%%%%%%%%%%%%%%%%%%%%%%%%%%%%%%%%%%%%%%%%%%%%%%%%
%% BACKMATTER
%%%%%%%%%%%%%%%%%%%%%%%%%%%%%%%%%%%%%%%%%%%%%%%%

\begin{theacknowledgments}
The simulations presented here were performed at the UK Astrophysical
Fluids Facility (UKAFF). 
\end{theacknowledgments}

%% If the bibliography is produced without BibTeX comment out the

\bibliographystyle{aipproc}   % if natbib is available

\bibliography{lodato}

\IfFileExists{\jobname.bbl}{}
 {\typeout{}
  \typeout{******************************************}
  \typeout{** Please run "bibtex \jobname" to optain}
  \typeout{** the bibliography and then re-run LaTeX}
  \typeout{** twice to fix the references!}
  \typeout{******************************************}
  \typeout{}
 }

\end{document}